\documentclass[a4paper,11pt]{article}
\usepackage[scaled]{DejaVuSansMono}
\usepackage[T1]{fontenc}
\usepackage[english]{babel}
\usepackage[nottoc]{tocbibind}
\usepackage{fontawesome}
\usepackage{amsmath}
\usepackage{twemojis}

\usepackage{microtype}
\usepackage[normalem]{ulem}
\usepackage[ruled, vlined,inoutnumbered,rightnl]{algorithm2e}
\SetAlFnt{\normalsize}
\SetAlCapFnt{\normalsize}
\SetAlCapNameFnt{\normalsize}

\usepackage[skins,listings]{tcolorbox}

\usepackage{xargs}
\usepackage[document]{ragged2e}
\usepackage{wrapfig}
\usepackage{svg}
\usepackage{lineno}
\usepackage{subcaption}
\usepackage[colorinlistoftodos,prependcaption]{todonotes}
\usepackage{empheq}
\usepackage{xcolor}

\usepackage{pos}

\definecolor{newblue}{HTML}{007fff}
\definecolor{awesome}{HTML}{FF2052}
\definecolor{sam}{RGB}{28, 28, 28}
\definecolor{myPink}{HTML}{EC407A}
\definecolor{myOrange}{HTML}{FFA726}
\definecolor{myGreen}{HTML}{66BB6A}
\definecolor{lfPurple}{HTML}{AB47BC}
\hypersetup{
  colorlinks=true,
  urlcolor=newblue,
  linkcolor=awesome,
  citecolor=lfPurple,
}

\usepackage{cleveref}
\usepackage{mathtools}

\makeatletter
\newcommand*\Acolorboxed[2][red]{%
   \let\bgroup{\romannumeral-`}%
   \@Acolorboxed{#1}#2&&\ENDDNE
}
\def\@Acolorboxed#1#2&#3&#4\ENDDNE{%
  \ifnum0=`{}\fi
  \setbox\z@\hbox{$\displaystyle#2{}\m@th$\kern\fboxsep \kern\fboxrule}%
  \edef\@tempa{\kern\wd\z@ \kern-\the\wd\z@ \fboxsep\the\fboxsep \fboxrule\the\fboxrule}%
  \@tempa
  \fcolorbox{#1}{#1}{\m@th$\displaystyle#2#3$}%
} 

\newcommand*\Aoutlinebox[2][red]{%
   \let\bgroup{\romannumeral-`}%
   \@Aoutlineboxed{#1}#2&&\ENDDNE
}
\def\@Aoutlinebox#1#2&#3&#4\ENDDNE{%
  \ifnum0=`{}\fi
  \setbox\z@\hbox{$\displaystyle#2{}\m@th$\kern\fboxsep \kern\fboxrule}%
  \edef\@tempa{\kern\wd\z@ \kern-\the\wd\z@ \fboxsep\the\fboxsep \fboxrule\the\fboxrule}%
  \@tempa
  \fcolorbox{#1}{white}{\m@th$\displaystyle#2#3$}%
} 

\newcommand*\Aoutlineboxed[2][red]{%
   \let\bgroup{\romannumeral-`}%
   \@Aoutlineboxed{#1}#2&&\ENDDNE
}
\def\@Aoutlineboxed#1#2&#3&#4\ENDDNE{%
  \ifnum0=`{}\fi
  \setbox\z@\hbox{$\displaystyle#2{}\m@th$\kern\fboxsep \kern\fboxrule}%
  \edef\@tempa{\kern\wd\z@ \kern-\the\wd\z@ \fboxsep\the\fboxsep \fboxrule\the\fboxrule}%
  \@tempa
  \fcolorbox{#1}{white}{\m@th$\displaystyle#2#3$}%
} 
\makeatother

\newcommandx{\fix}[2][1=]{\todo[linecolor=red,backgroundcolor=red!25,bordercolor=red,#1]{#2}}

\newcommandx{\thiswillnotshow}[2][1=]{\todo[disable,#1]{#2}}

\title{MLMC: Machine Learning Monte Carlo for Lattice Gauge Theory}

\author*[a]{Sam Foreman}
\author[a,b]{Xiao-Yong Jin}
\author[a,b]{James C. Osborn}

\affiliation[a]{Leadership Computing Facility, Argonne National Laboratory,\\
  9700 S. Cass Ave, Lemont IL, USA}

\affiliation[b]{Computational Science Division, Argonne National Laboratory,\\
9700 S. Cass Ave, Lemont IL, USA}

\emailAdd{foremans@anl.gov}
\emailAdd{xjin@anl.gov}
\emailAdd{osborn@alcf.anl.gov}

\abstract{%
We present a trainable framework for efficiently generating gauge
configurations, and discuss ongoing work in this direction.
In particular, we consider the problem of sampling configurations from a 4D $SU(3)$ lattice gauge theory, and consider a generalized leapfrog integrator in the molecular dynamics update that can be trained to improve sampling efficiency.
Code is available online at \href{https://github.com/saforem2/l2hmc-qcd}{\faGithub \texttt{l2hmc-qcd}}.
}

\FullConference{The 40th International Symposium on Lattice Field Theory (Lattice 2023)\\
July 31st - August 4th, 2023\\
Fermi National Accelerator Laboratory\\}

\begin{document}
\maketitle

\tableofcontents

\section{Introduction}
\label{sec:background}

We would like to calculate observables $\mathcal{O}$:
\begin{equation}
\left\langle \mathcal{O}\right\rangle \propto \int \left[\mathcal{D} x\right]\, \mathcal{O}(x)\, \pi(x)
\end{equation}
where $\pi(x) \propto e^{-\beta S(x)}$ is our target distribution.
If these were independent, we could approximate the integral as
$\left\langle\mathcal{O}\right\rangle \simeq \frac{1}{N}\sum_{n=1}^{N}
\mathcal{O}(x_{n})$ with variance
\begin{equation}
\sigma_{\mathcal{O}}^{2} = \frac{1}{N}\,\mathrm{Var}\left[\mathcal{O}(x)\right] \Longrightarrow \sigma_{\mathcal{O}} \propto \frac{1}{\sqrt{N}}.
\end{equation}
Instead, nearby configurations are correlated, causing us to incur a factor of
$\tau_{\mathrm{int}}^{\mathcal{O}}$ in the variance expression
\begin{equation}
\sigma_{\mathcal{O}}^{2} = \frac{\tau_{\mathrm{int}}^{\mathcal{O}}}{N} \mathrm{Var}\left[\mathcal{O}(x)\right].
\end{equation}

\subsection{Hamiltonian Monte Carlo (HMC)}
\label{subsec:hmc}
The typical approach~\cite{foreman_deep_2021,foreman_leapfroglayers_2022} is to use 
Hamiltonian Monte Carlo (HMC) algorithm for generating configurations 
distributed according to our target distribution $\pi(x)$.
This can be done by sequentially constructing a chain of states $\{x_{0},\, x_{1},\, x_{2},\, \ldots,\, x_{i},\, \ldots,\, x_{n}\}$, such that, as $n \rightarrow \infty$:
%
%
\begin{wrapfigure}[17]{l}{0.37\textwidth}
  \begin{center}
  \caption{\label{fig:hmc-update}Leapfrog update.}
  \includegraphics[width=\linewidth]{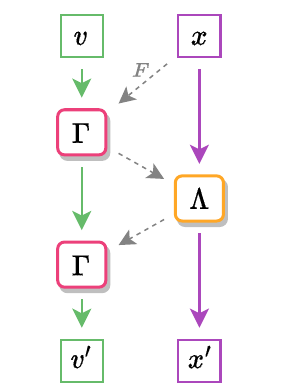}
  \end{center}
\end{wrapfigure}
%
%
\begin{equation}
\left\{x_{i}, x_{i+1}, x_{i+2}, \ldots, x_{n}\right\} \sim \pi(x).
\end{equation}
To do this, we begin by introducing a fictitious momentum\footnote{Here $\sim$
means \textit{is distributed according to}.} $v \sim \mathcal{N}(0, 1)$
normally distributed, independent of $x$.
We can write the joint distribution $\pi(x, v)$ as
\begin{align}
\pi(x, v) &= \pi(x) \pi(v) \propto e^{-S(x)} e^{-\frac{1}{2} v^{T}v} \\
&= e^{-\left[S(x) + \frac{1}{2} v^{T} v \right]}
\end{align}
We can evolve the Hamiltonian dynamics of the $(\dot{x}, \dot{v}) =
(\partial_{v} H, -\partial_{x} H)$ system using operators $\Gamma: v
\rightarrow v'$ and $\Lambda: x \rightarrow x'$.
Explicitly, for a single update step of the leapfrog integrator:
\begin{align}
\tilde{v} &\coloneqq \Gamma(x, v) = v - \frac{\varepsilon}{2} F(x) \\
x' &\coloneqq \Lambda(x, \tilde{v}) = x + \varepsilon \tilde{v} \\
v' &\coloneqq \Lambda(x', \tilde{v}) = \tilde{v} - \frac{\varepsilon}{2} F(x'),
\end{align}
where we've written the force term as $F(x) = \partial_{x}S(x)$.
Typically, we build a trajectory of $N_{\mathrm{LF}}$ leapfrog steps $(x_{0}, v_{0}) \rightarrow (x_{1}, v_{1}) \rightarrow \cdots \rightarrow (x', v'),$ and propose $x'$ as the next state in our chain.
%
%
This proposal state is then accepted according to the Metropolis-Hastings
criteria~\cite{robert_metropolis-hastings_2016}
\begin{equation}
A(x'|x) = \mathrm{min}\left\{{1, \frac{\pi(x')}{\pi(x)} \left| \frac{\partial x'}{\partial x} \right|}\right\}.
\end{equation}
\section{Method}
\label{sec:method}
\begin{wrapfigure}[19]{r}{0.42\textwidth}
    \begin{center}
        \caption{\label{fig:lf-layer}Generalized MD update.} 
        \includegraphics[width=\linewidth]{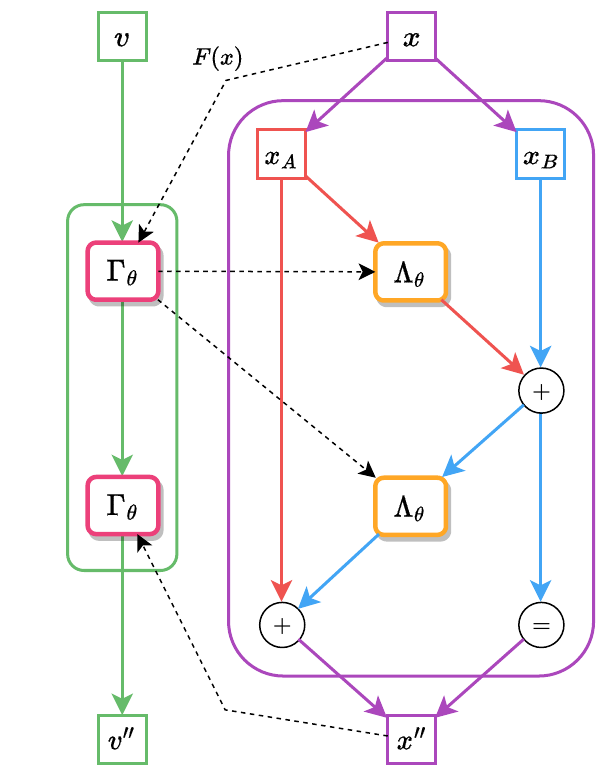}
  \end{center}
\end{wrapfigure}
Unfortunately, HMC is known to suffer from long auto-correlations and often
struggles with multi-modal target densities.
To combat this, we propose building on the approach from~\cite{foreman_learning_2019,foreman_deep_2021,foreman_leapfroglayers_2022}. 
We introduce two (invertible) neural networks $\texttt{xNet}: (x, v) \rightarrow (\alpha_{x}, \beta_{x}, \gamma_{x})$, $\texttt{vNet}: (x, F) \rightarrow (\alpha_{v}, \beta_{v}, \gamma_{v})$.
%
%

Here, $\left(\alpha, \beta, \gamma\right)$ are all of the same dimensionality as $x$ and $v$, and are parameterized by a set of weights $\theta$.
These network outputs $(\alpha, \beta, \gamma)$ are then used in a generalized MD update (as
shown in Fig~\ref{fig:lf-layer}) via:
\begin{align}
\Aoutlineboxed[myPink]{\Gamma^{\pm}_{\theta}&}: (x, v) \rightarrow (x, v'), \\
\Aoutlineboxed[myOrange]{\Lambda^{\pm}_{\theta}&}: (x, v) \rightarrow (x', v).
\end{align}
where the superscript $\pm$ on $\Gamma^{\pm}_{\theta}$, $\Lambda^{\pm}_{\theta}$
correspond to the direction $d \sim \mathcal{U}(-1, +1)$ of the update.
%

%
To ensure that our proposed update remains reversible, we split the $x$ update into two sub-updates on complementary subsets ($x = x_{A} \cup x_{B}$):
\begin{align}
    {v'} &= {\Gamma^{\pm}_{\theta}(x, v)} \\
    {x'} &= x_{B} + {\Lambda^{\pm}_{\theta}(x_{A}, v')} \\
    {x''} &= x'_{A} + {\Lambda^{\pm}_{\theta}(x'_{B}, v')} \\
    {v''} &= {\Gamma^{\pm}_{\theta}(x'', v')}
\end{align}
%
%
\subsection{\label{subsec:algorithm}Algorithm}
\begin{enumerate}
    \item \texttt{input:} $x$
    \begin{itemize}
        \item Re-sample $v \sim \mathcal{N}(0, 1)$
        \item Construct initial state $\xi \coloneqq (x, v)$
    \end{itemize}
    \item \texttt{forward:} Generate proposal $\xi'$ by passing initial $\xi$
      through $N_{\mathrm{LF}}$ leapfrog layers:
    \begin{equation}
        \xi \xrightarrow[]{\mathrm{LF\,\,\, Layer}} \xi_{1} \rightarrow \cdots \rightarrow \xi_{N_{\mathrm{LF}}} = \xi' \coloneqq (x'', v'')
    \end{equation}
    \begin{itemize}
        \item Metropolis-Hastings accept / reject:
        \begin{equation}
            A(\xi'|\xi) = \mathrm{min}\left\{1, \frac{\pi(\xi')}{\pi(\xi)} \left|\mathcal{J}\left(\xi', \xi\right)\right|\right\},
            \label{eq:MH}
        \end{equation}
        where $\left|\mathcal{J}(\xi',\xi)\right|$ is the determinant of the
        Jacobian.
    \end{itemize}
    \item \texttt{backward:} (if training)
    \begin{itemize}
        \item Evaluate the loss function $\mathcal{L}(\xi', \xi)$ and back
          propagate
    \end{itemize}
    \item \texttt{return:} $x_{i+1}$
    \begin{itemize}
        \item Evaluate MH criteria (Eq.~\ref{eq:MH}) and return accepted
          config:
        \begin{equation}
            x_{i+1} \gets \begin{cases}
                x'' \quad \text{w/ prob.}\quad A(\xi'|\xi)  \\
                x \,\,\,\quad \text{w/ prob.} \quad 1 - A(\xi'|\xi)
            \end{cases}
        \end{equation}
    \end{itemize}
\end{enumerate}
\section{\label{sec:lgt}Lattice Gauge Theories}
\begin{wrapfigure}[21]{R}{0.42\textwidth}
  \begin{center}
  \caption{\label{fig:csd}$\delta Q \rightarrow 0$ with increasing $\beta$ for the 2D $U(1)$ model. Image from~\cite{foreman_leapfroglayers_2022}.}
  \includegraphics[width=\linewidth]{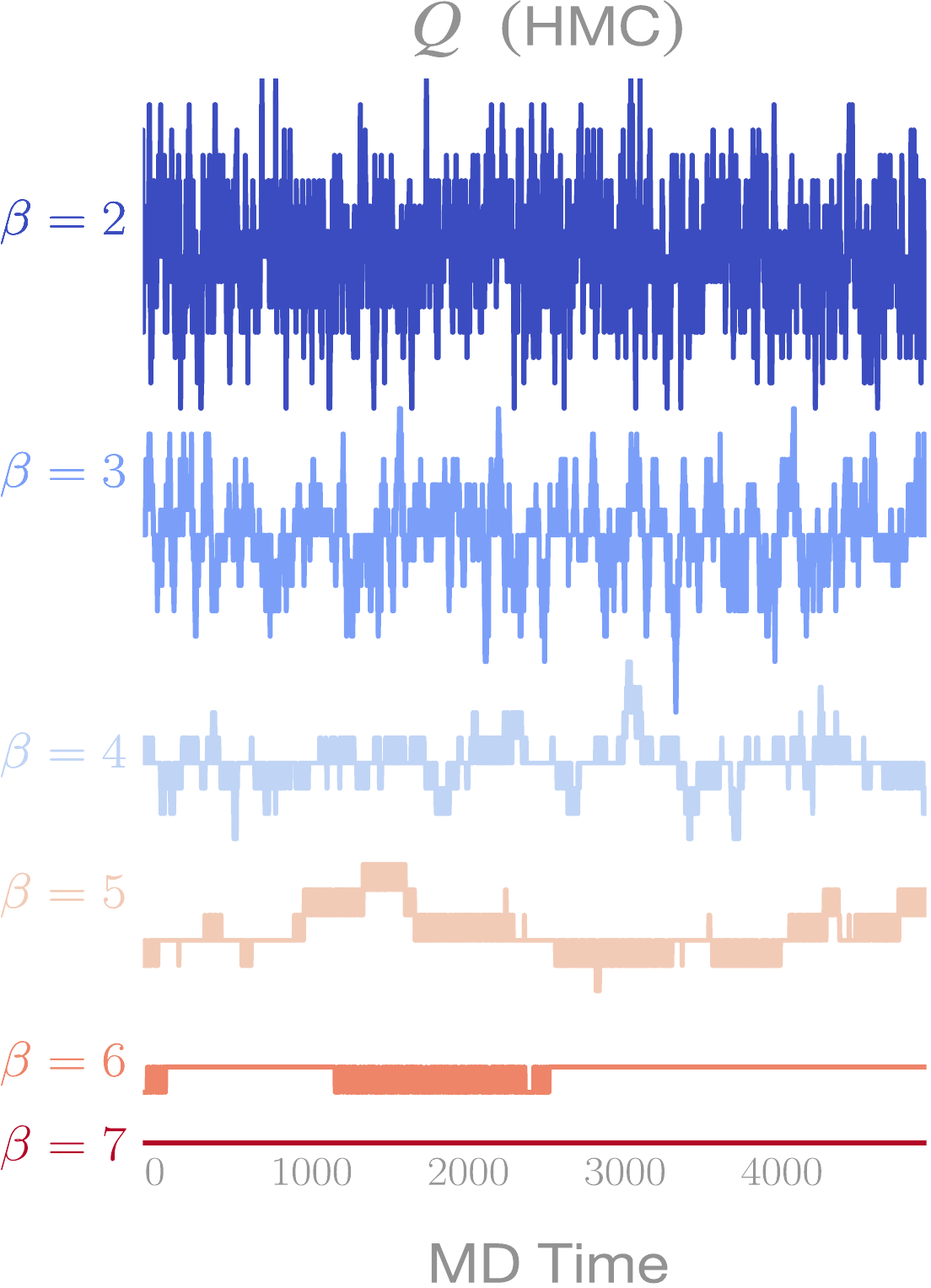}
  \end{center}
\end{wrapfigure}
\subsection{\label{subsec:2dU1}2D \texorpdfstring{$U(1)$}{U(1)} Model}
We build upon the approach originally introduced in~\cite{levy_generalizing_2018}, which was successfully applied to the 2D $U(1)$ lattice gauge model in~\cite{foreman_learning_2019,foreman_deep_2021,foreman_leapfroglayers_2022}.
In particular, we are interested in measuring the (scalar) topological charge
$Q \in \mathbb{Z}$ on the lattice.
Since different lattice configurations with the same value of $Q$ are related
by a gauge transformation, they do not meaningfully contribute to our
statistics.

Because of this, we would like to generate configurations from different
\textit{topological sectors} (characterized by different values of $Q$)
to reduce uncertainty in our statistical estimates.
By repeating this procedure at increasing spatial resolution\footnote{Here $a$ is the lattice spacing.} ($\beta \propto 1
/ a$), we are able to extrapolate our
estimates to the continuum limit where they can be compared with experimental
measurements.
Current approaches such as HMC are known to suffer from auto-correlation times
which scale exponentially in this limit, significantly limiting their
effectiveness.
This phenomenon can be seen in Fig~\ref{fig:csd}, where fluctuations in the
topological charge between sequential configurations (the \emph{tunneling rate}) $\delta Q = |Q^{i + 1} - Q^{i}|$ decreases as $\beta= 2
\rightarrow 3 \rightarrow \cdots$, and disappear completely ($Q =
\mathrm{const.}$) by $\beta = 7$.
%
%

\subsubsection{\label{subsubsec:2dResults}Results}
\begin{figure}[htpb!]
    \centering
    \hfill
    \begin{subfigure}{0.4\textwidth}
        \includegraphics[width=\textwidth]{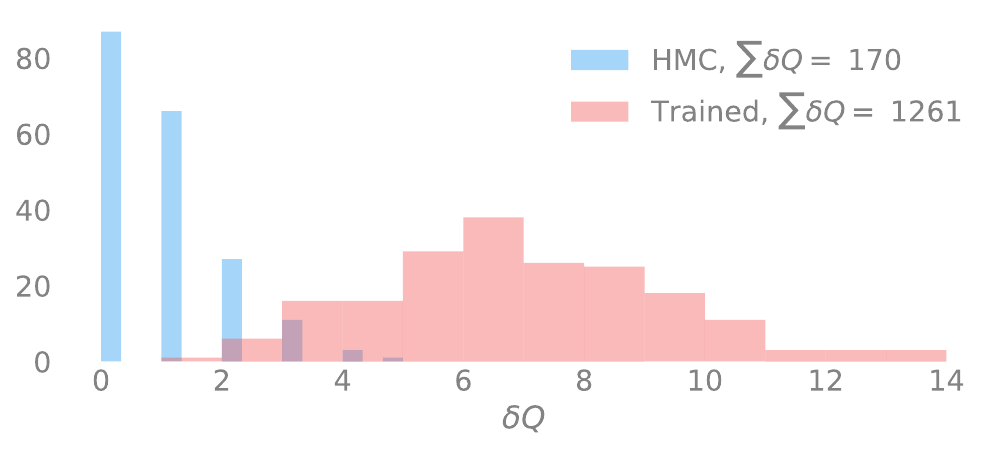}
        \caption{\label{subfig:dqhist}$\delta Q$ for trained model (red) vs HMC (blue).}
    \end{subfigure}
    \hfill
    \begin{subfigure}{0.55\textwidth}
        \includegraphics[width=\textwidth]{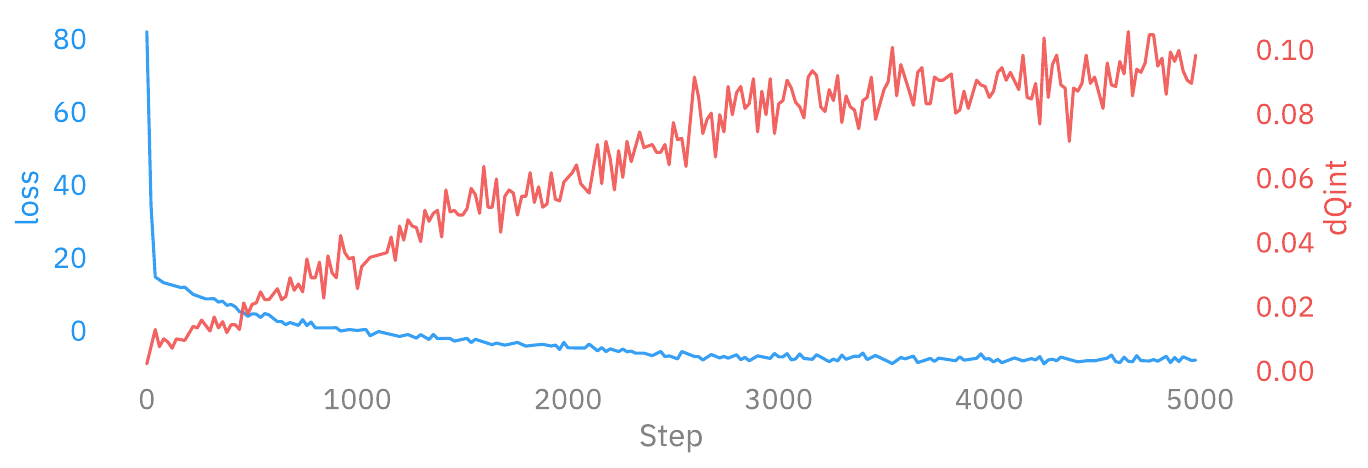}
        \caption{\label{subfig:loss_dQint}Loss (blue) and $\delta Q$ (red) during training}
    \end{subfigure}
    \hfill
    \begin{subfigure}{0.30\textwidth}
        \includegraphics[width=\textwidth]{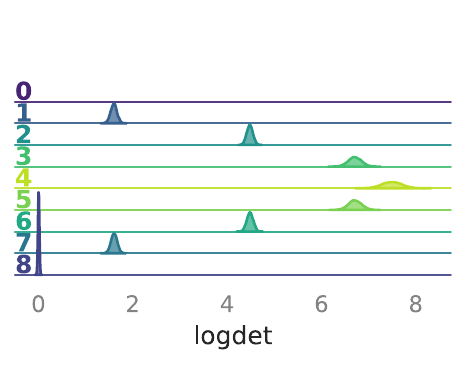}
        \caption{\label{subfig:logdet2dU1}$|\mathcal{J}|$ vs LF step \emph{(trained)}}
    \end{subfigure}
    \hfill
    \begin{subfigure}{0.30\textwidth}
        \includegraphics[width=\textwidth]{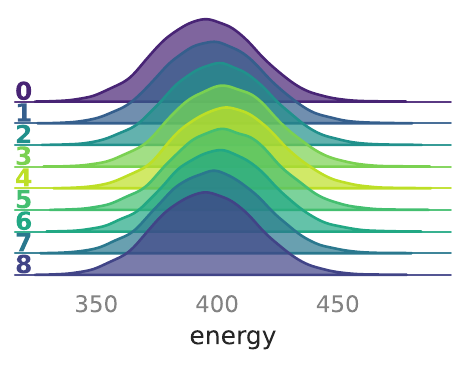}
        \caption{\label{subfig:energy2dU1}$H$ vs LF step \emph{(trained)}}
    \end{subfigure}
    \hfill
    \begin{subfigure}{0.30\textwidth}
        \includegraphics[width=\textwidth]{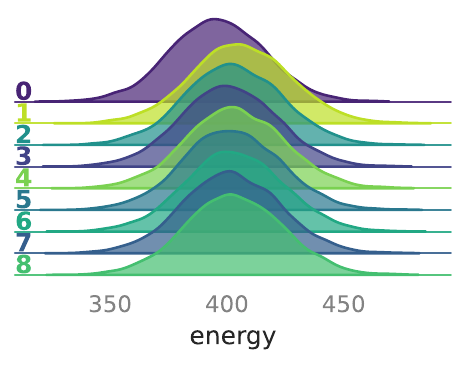}
        \caption{\label{subfig:energy2dU1hmc}$H$ vs LF step \emph{(HMC)}}
    \end{subfigure}
    \hfill
    \caption{\label{fig:2dU1}Results from trained 2D $U(1)$ model at $\beta = 4$. In~\ref{subfig:energy2dU1} we see the energy $H$ increasing towards the middle of the trajectory, resulting in improved tunneling rate (larger $\delta Q$) in ~\ref{subfig:dqhist}. \href{https://saforem2.github.io/l2hmc-qcd/qmd/l2hmc-2dU1/l2hmc-2dU1.html}{\twemoji{blue book} Jupyter notebook}.}
\end{figure}

Results for the 2D $U(1)$ model trained at $\beta = 4$ in $\simeq 25$ minutes on a single NVIDIA A100 GPU, using \href{https://github.com/saforem2/l2hmc-qcd}{\faGithubAlt \texttt{l2hmc-qcd}}.
We provide the full \href{https://saforem2.github.io/l2hmc-qcd/qmd/l2hmc-2dU1/l2hmc-2dU1.html}{\twemoji{blue book} Jupyter notebook} containing the results in Fig~\ref{fig:2dU1}.
%
%
%
\subsection{\label{subsec:4dSU3}4D \texorpdfstring{$SU(3)$}{SU(3)} Model}
%
%
We would like to generalize this approach to handle 4D $SU(3)$ link variables $U_{\mu}(n) \in SU(3)$:
\begin{equation}
U_{\mu}(n) = \exp\left[i \omega_{\mu}^{k}(n) \lambda^{k} \right]
\end{equation}
where $\omega_{\mu}^{k}(n) \in \mathbb{R}$ and $\lambda^{k}$ are the generators
of $SU(3)$.
We consider the standard Wilson gauge action
\begin{align}
S_{G} &= -\frac{\beta}{6}\sum \mathrm{Tr}\left[U_{\mu\nu}(n) + U^{\dagger}_{\mu\nu}(n) \right],\quad\text{where} \\
U_{\mu\nu}(n) &= U_{\mu}(n) U_{\nu}(n + \hat{\mu}) U_{\mu}^{\dagger}(n + \hat{\nu}) U^{\dagger}_{\nu}(n).
\end{align}
\subsubsection{\label{subsubsec:hmc-4dSU3}Generic MD Updates}
As before, we introduce momenta $P_{\mu}(n) = P^{k}_{\mu}(n) \lambda^{k}$
conjugate to the real fields $\omega_{\mu}^{k}(n)$.
We can write the Hamiltonian as
\begin{equation}
    H[P, U] = \frac{1}{2} P^{2} + S_{G}[U] \Longrightarrow
    \Aoutlinebox[myGreen]{\frac{d\omega^{k}}{dt} = \frac{\partial H}{\partial P^{k}}},
    \quad \Aoutlinebox[lfPurple]{\frac{dP^{k}}{dt} = - \frac{\partial H}{\partial \omega^{k}}}.
\end{equation}
%
%
%
To update the gauge field $U_{\mu} = e^{i\omega_{\mu}^{k} \lambda^{k}}$, write
$\Aoutlinebox[myGreen]{\frac{d\omega^{k}}{dt}\lambda^{k} = P^{k} \lambda^{k}}$
and discretize with step size $\varepsilon$:
\begin{align}
    -i \log U(\varepsilon) &= - i \log U(0) + \varepsilon P(0) \\
    U(\varepsilon) &= e^{i \varepsilon P(0)} U(0) \Longrightarrow \\
    \Aoutlineboxed[myOrange]{\Lambda: U \rightarrow U'} &= e^{i \varepsilon P} U.
\end{align}
Similarly for the momentum update $\Aoutlineboxed[lfPurple]{\frac{dP^{k}}{dt} = - \frac{\partial H}{\partial \omega^{k}}}$,
\begin{align}
    P(\varepsilon) &= P(0) - \varepsilon F[U]\\
    \Aoutlineboxed[myPink]{\Gamma: P \rightarrow P'} &= P - \frac{\varepsilon}{2} F[U]
\end{align}
where $F[U]$ is the force term (see~\ref{subsec:force_term}).
%
%
%
\subsubsection{\label{subsubsec:generalized-md-4dSU3}Generalized MD Update}
As in Sec.\ref{sec:method}, we introduce \texttt{pNet}: $(U, F) \rightarrow \left(\alpha_{P}, \beta_{P},
\gamma_{P}\right)$ and \texttt{uNet}: $(U, P) \rightarrow \left(\,\cdot\,, \beta_{U}, \gamma_{U}\right)$. Note that we have omitted the $U$ scaling term ($\alpha_{U}$) term in this update since $U \in SU(3)$.
In terms of the generalized update operators,
\begin{align}
    \Aoutlineboxed[myPink]{\Gamma^{\pm}_{\theta}}&: (U, P) \xrightarrow[]{\left(\alpha_{P}, \beta_{P}, \gamma_{P}\right)} (U, P') \\
    \Aoutlineboxed[myOrange]{\Lambda^{\pm}_{\theta}}&: (U, P)\xrightarrow[]{\left(\, \cdot ,\, \beta_{U}, \gamma_{U}\right)} (U', P)
\end{align}
we can write the complete update:
\begin{align}
    P' &= \Gamma^{\pm}_{\theta}(U, P) \\
    U' &= U_{B} + \Lambda^{\pm}_{\theta}(U_{A}, P') \\ 
    U'' &= U'_{A} + \Lambda^{\pm}_{\theta}(U'_{B}, P')\\
    P'' &= \Gamma^{\pm}_{\theta}(U'', P')
\end{align}
%

\textbf{Momentum Update}\\
In this case, our $\texttt{pNet}: (U, F) = (\alpha_{P},
\beta_{P}, \gamma_{P})$.
We can write the generalized momentum update as $P^{\pm} \coloneqq \Gamma_{\theta}^{\pm}(U, P)$, where\footnote{Note that $\left(\Gamma^{+}\right)^{-1} = \Gamma^{-}$, i.e. $\Gamma^{+}\left[\Gamma^{-}(U, F)\right] = \Gamma^{-}\left[\Gamma^{+}(U, F)\right] = (U, F)$, and similarly for $\Lambda^{\pm}$}:
%
%
%
\begin{enumerate}
    \item \texttt{forward}, $(+)$:
    \begin{equation}
        P^{+} \coloneqq \Gamma_{\theta}^{+}(U, P) = P \cdot e^{\frac{\varepsilon}{2} \alpha_{P}} - \frac{\varepsilon}{2}\left[ F \cdot e^{\varepsilon \beta_{P}} + \gamma_{P} \right]
    \end{equation}
    \item \texttt{backward}, $(-)$:
    \begin{equation}
        P^{-} \coloneqq \Gamma_{\theta}^{-}(U, P) = e^{-\frac{\varepsilon}{2} \alpha_{P}} \cdot \left\{P + \frac{\varepsilon}{2} \left[ F \cdot e^{\varepsilon \beta_{P}} + \gamma_{P} \right] \right\}.
    \end{equation}
\end{enumerate}
By introducing the above modifications, we incur a factor of 
$\log\left|\frac{\partial P^{\pm}}{\partial P}\right| = \, \pm \, \frac{\varepsilon}{2}\sum \alpha_{P}$
in the Metropolis Hastings accept / reject $A(U'|U)$, and the sum is taken over the full trajectory.\\
%
\textbf{Link Update}\\
Similarly to the momentum update, the outputs from our 
$\texttt{uNet}: (U, P) \rightarrow \left(\, \cdot\,, \beta_{U}, \gamma_{U}\right)$
are used in the generalized link update $U^{\pm} \coloneqq \Lambda_{\theta}^{\pm}(U, P) = e^{i\varepsilon \tilde{P}^{\pm}} U$
%
%
(where $\tilde{P}^{\pm} \in \mathfrak{su(3)}$).
Explicitly:
\begin{enumerate}
\item \texttt{forward}, $(+)$:
    \begin{equation}
        U^{+} \coloneqq \Lambda^{+}_{\theta}(U, P) = e^{i\varepsilon \tilde{P}^{+}} U, \quad \text{with}\quad \tilde{P}^{+} = \left[P\cdot e^{\varepsilon \beta_{U}} + \gamma_{U}\right]
    \end{equation}
\item \texttt{backward}, $(-)$:
    \begin{equation}
        U^{-} \coloneqq \Lambda^{-}_{\theta}(U, P) = e^{i\varepsilon \tilde{P}^{-}} U, \quad \text{with}\quad \tilde{P}^{-} = e^{-\varepsilon \beta_{U}} \cdot \left[P - \gamma_{U} \right] 
    \end{equation}
\end{enumerate}
\subsection{\label{subsec:training}Training}
We construct a loss function using the expected squared charge
difference
\begin{equation}
    \mathcal{L}_{\theta}(U, U') = \mathbb{E}\left[A(U'|U)\,\cdot \delta^{2}_{Q}(U, U')\right],
\end{equation}
where $\delta^{2}_{Q}(U, U')= |Q' - Q|^{2}$ is the squared topological
charge (see~\ref{subsec:topological_charge}) difference between the initial and proposal configurations.

\subsection{\label{sec:results}Results}
\begin{figure}[htpb!]
    \centering
    \begin{subfigure}{0.31\textwidth}
        \includegraphics[width=\textwidth]{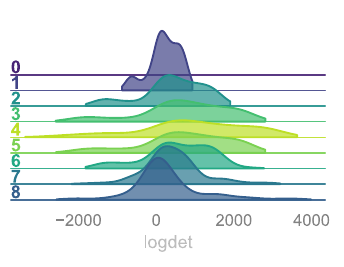}
        \caption{\label{subfig:logdet1} 100 train steps}
    \end{subfigure}
    \hfill
    \begin{subfigure}{0.31\textwidth}
        \includegraphics[width=\textwidth]{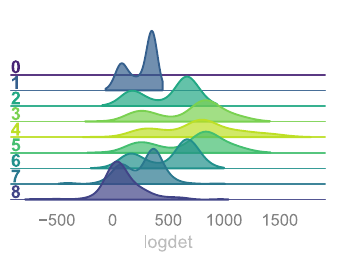}
        \caption{\label{subfig:logdet2} 500 train steps}
    \end{subfigure}
    \hfill
    \begin{subfigure}{0.31\textwidth}
        \includegraphics[width=\textwidth]{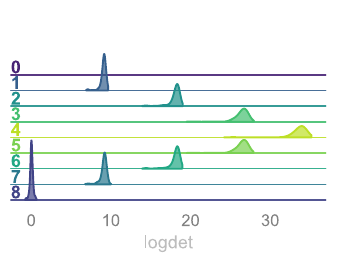}
        \caption{\label{subfig:logdet3} 1000 train steps}
    \end{subfigure}
    \caption{\label{fig:logdet}Evolution of $\left|\mathcal{J}\right|$ during the first 1000 training iterations for the 4D $SU(3)$ model.}
\end{figure}

For the trained 2D $U(1)$ model (Fig~\ref{fig:2dU1}), we see in Fig~\ref{subfig:logdet2dU1} that 
$\left|\mathcal{J}\right|$
increases towards the middle of the trajectory, allowing for the sampler to overcome the large energy barriers between different topological sectors.
This results in a greater \emph{tunneling rate} ($\delta Q$) when compared to generic HMC.
Identical behavior is observed after a short training run for the 4D $SU(3)$ model, as shown in Fig~\ref{fig:logdet}.
%

\section{\label{sec:conclusion}Conclusion}
In this work we've introduced a generalized MD update for generating 4D $SU(3)$ gauge configurations that can be trained to improve sampling efficiency.
Note that this is a relatively simple proof of concept demonstrating how to construct such a sampler.
In a future work we plan to further investigate (and quantify) the cost / benefit when compared to alternative approaches such as traditional HMC and purely generative (OT / KL-Divergence~\cite{albergo_flow-based_2019,albergo_introduction_2021,boyda_sampling_2021,kanwar_equivariant_2020}) based approaches.
%
%
\section{\label{sec:acknowledgements}Acknowledgements}
This research used resources of the Argonne Leadership Computing Facility,
which is a DOE Office of Science User Facility supported under Contract DE-AC02-06CH11357.
This research was supported by the Exascale Computing Project (17-SC-20-SC), a collaborative effort of the U.S. Department of Energy Office of Science and the National Nuclear Security Administration.
%
%
\nocite{*}
\bibliography{references}{}
\bibliographystyle{abbrvnat}


\appendix
\section{\label{appendix}Appendix}
%
%
\subsection{\label{subsec:force_term}Force Term}

We can write the force term as
\begin{equation}
    F = - \frac{1}{\lambda^{2}} \sum_{k} \lambda^{k} \, \mathrm{Tr}\left[ i \left(UA - A^{\dagger}U^{\dagger}\right) \lambda^{k} \right]
\end{equation}
where $A$ is the sum over staples
\begin{align}
A = \sum_{\mu\neq\nu}& U_{\mu}(x +\hat{\mu})\, U^{\dagger}_{\mu}(x + \hat{\nu}) \, U^{\dagger}_{\nu}(x) \\ 
&+ \sum_{\mu\neq\nu} U_{-\nu}(x + \hat{\mu})\, U^{\dagger}_{\mu}(x - \hat{\nu})\, U^{\dagger}_{-\nu}(x).
\end{align}
Since, $i\left(UA - A^{\dagger}U^{\dagger}\right) \in \mathfrak{su}(3)$,
we can write it in terms of the generators $\lambda^{k}$ as
\begin{align}
    \sum_{k} \lambda^{k} \,\mathrm{Tr}\left[ \lambda^{k} \sum_{j} c_{j}\, \lambda^{j}\right] &= \sum_{k}\sum_{j} c_{j}\, \lambda^{j} \,\mathrm{Tr}\left[ \lambda^{k} \, \lambda^{j}\right] \\
    &= \frac{1}{2}\sum_{k}\sum_{j} c_{j}\, t^{k}\, \delta_{jk} \\
    &= \frac{1}{2}\sum_{k} c_{k} \, t^{k}
\end{align}
consequently, we can simplify the force term as
\begin{equation}
    F[U] = - \frac{1}{2g^{2}}\, i \, \left(UA - A^{\dagger}U^{\dagger}\right).
\end{equation}

\subsection{\label{subsec:topological_charge}Topological Charge \texorpdfstring{$Q$}{Q}}
In lattice field theory, the topological charge $Q$ is defined as the 4D
integral over spacetime of the topological charge density $q$.
In the continuum,
\begin{align}
    Q &= \int d^{4}x q(x), \text{ where } \\
    q(x) &= \frac{1}{32\pi^{2}} \epsilon_{\mu\nu\rho\lambda} \mathrm{Tr}\left\{ F_{\mu\nu} F_{\rho\lambda} \right\}
\end{align}
On the lattice, we choose a discretization\footnote{We are free to choose a specific discretization as long as it gives the right continuum limit} $q_{L}(x)$ such that
$Q = a^{4} \sum_{x} q_{L}(x)$.
The most obvious discretization of $q_{L}$ uses the $1\times1$ plaquette
$P_{\mu\nu}(x)$, and can be written as
\begin{equation}
    q^{\mathrm{plaq}}_{L}(x) = \frac{1}{32\pi^{2}} \epsilon_{\mu\nu\rho\lambda} \mathrm{Tr}\left\{P_{\mu\nu}(x) P_{\rho\lambda}(x)\right\}
\end{equation}
this has the advantage of being computationally inexpensive, but leads
to lattice artifacts of order $\mathcal{O}(a^{2})$.

\end{document}